\documentclass[prb,longbibliography,showpacs, showkeys, twocolumn,superscriptaddress,amsmath,amssymb,verbatim]{revtex4-2}

\usepackage{graphicx}
\usepackage{lmodern}
\usepackage{epstopdf}
\usepackage{dcolumn}
\usepackage{bm}
\usepackage{subfigure}
\usepackage{amsmath} 

\usepackage[pdftex,colorlinks=true,citecolor=blue,linkcolor=blue,urlcolor=blue,bookmarks=true]{hyperref}
\usepackage[sort&compress]{natbib}

\usepackage{color}
\usepackage{textcomp}
\definecolor{red}{rgb}{1,0,0}

\definecolor{blue}{rgb}{0,0,1}

\definecolor{green}{rgb}{0,1,0}

\definecolor{lblue}{rgb}{0.4, 0.8, 0.8}

\begin{document}
\preprint{APS}

\title{Incommensurate magnetic order arising from frustrated interchain interactions in the spin-1/2 chain compound AgCuVO$_4$}

\author{A. Hromov}
\affiliation{Jo\v{z}ef Stefan Institute, Jamova cesta 39, 1000 Ljubljana, Slovenia}
\affiliation{Faculty of Mathematics and Physics, University of Ljubljana, Jadranska ulica 19, 1000 Ljubljana, Slovenia}
\author{A. Zorko}
\affiliation{Jo\v{z}ef Stefan Institute, Jamova cesta 39, 1000 Ljubljana, Slovenia}
\affiliation{Faculty of Mathematics and Physics, University of Ljubljana, Jadranska ulica 19, 1000 Ljubljana, Slovenia}
\author{M. Gomil\v{s}ek}
\affiliation{Jo\v{z}ef Stefan Institute, Jamova cesta 39, 1000 Ljubljana, Slovenia}
\affiliation{Faculty of Mathematics and Physics, University of Ljubljana, Jadranska ulica 19, 1000 Ljubljana, Slovenia}
\author{I. Puente Orench}
\affiliation{Institut Laue-Langevin, BP 156, 38042 Grenoble Cedex 9, France}
\author{L. Keller}
\affiliation{PSI Center for Neutron and Muon Sciences, 5232 Villigen PSI, Switzerland}
\author{T. Shiroka}
\affiliation{PSI Center for Neutron and Muon Sciences, 5232 Villigen PSI, Switzerland}
\affiliation{Laboratorium für Festkörperphysik, ETH Zürich, Zürich, CH-8093, Switzerland}
\author{A. Prokofiev}
\affiliation{Institute of Solid State Physics, Vienna University of Technology, Wiedner Hauptstrasse 8-10, 1040 Vienna, Austria}
\author{M. Pregelj}
\email{matej.pregelj@ijs.si}
\affiliation{Jo\v{z}ef Stefan Institute, Jamova cesta 39, 1000 Ljubljana, Slovenia}
\affiliation{Faculty of Mathematics and Physics, University of Ljubljana, Jadranska ulica 19, 1000 Ljubljana, Slovenia}

\date{\today}

\begin{abstract}

Quantum spin chains with competing interactions offer a platform where low dimensionality and frustration—both acting to suppress magnetic order—intersect.
We studied magnetic ordering in the spin-1/2 chain compound AgCuVO$_4$ using muon spin spectroscopy and neutron diffraction.
Long-range magnetic order emerges at $T_N = 2.0(1)$\,K, which is $\sim$1/200 of the dominant intrachain coupling $J$ and $\sim$1/15 of the interchain interactions.
The collinear incommensurate amplitude-modulated magnetic structure features a reduced ordered moment of 0.13(2)\,$\mu_\mathrm{B}$, confined to the $ab$ plane and modulated along the $c$ axis—perpendicular to the spin chains—indicating frustrated interchain couplings.
The low $T_N$, small moment, and incommensurate order highlight strong frustration, positioning AgCuVO$_4$ as a model system for exploring frustration in quantum spin chains.

\end{abstract}

\keywords{spin chain, magnetic order, frustration}
\maketitle

\section{Introduction}

Quantum spin systems with reduced dimensionality exhibit unusual and often counterintuitive phenomena \cite{lacroix2011introduction, khatua2023experimental}, such as disordered, yet highly-entangled, spin-liquid states where fractional spinon excitations give rise to a broad, continuum-like excitation spectrum instead of a sharp magnon-like dispersion \cite{mourigal2013fractional,pregelj2018coexisting,janvsa2018observation}. 
A prototypical example is the spin-1/2 chain, where magnetic order is suppressed irrespective of the strength of nearest-neighbor interactions \cite{vasiliev2018milestones,haldane1985theta}.
In real materials, however, weak but finite interchain couplings are always present, and these typically stabilize a long-range magnetic order at sufficiently low temperatures \cite{vasiliev2018milestones,Katanin_2007}.
Yet, magnetic frustration—arising from competing interactions, such as next-nearest-neighbor antiferromagnetic intrachain couplings—can also inhibit magnetic ordering, even in the presence of interchain couplings, leading to spin-liquid behavior driven by geometrical constraints \cite{lacroix2011introduction}.
Spin chains with such competing interactions thus represent a particularly compelling model system, where low dimensionality and frustration—two powerful mechanisms that suppress magnetic order—converge.

Well-known spin-chain systems are cuprates, where chains typically consist of CuO$_4$ plaquettes.
The nature of the exchange interaction depends on the Cu--O--Cu bond geometry: edge-sharing CuO$_4$ plaquettes yield bond angles near 90$^\circ$ promoting ferromagnetic exchange interactions between spin-1/2 Cu$^{2+}$ magnetic ions, while corner-sharing plaquettes configuration approach 180$^\circ$ bond angles favoring antiferromagnetic exchange interactions.
For example, Sr$_2$CuO$_3$ exhibits nearly ideal antiferromagnetic spin-1/2 chains with a strong 180$^\circ$ exchange path and an interaction strength of $\sim$2000 K \cite{motoyama1996magnetic}.
In contrast, LiCuVO$_4$ features edge-sharing plaquettes \cite{enderle2005quantum}, resulting in frustrated chains with ferromagnetic nearest-neighbor and antiferromagnetic next-nearest-neighbor interactions on the order of tens of kelvin.
Compounds with Cu--O--Cu angles near the crossover between ferro- and antiferromagnetic regimes are relatively rare, yet particularly intriguing.
In such systems, the weakened intrachain exchange interactions may enhance the role of interchain couplings and other perturbative effects, potentially giving rise to novel and complex magnetic behavior.

\begin{figure*}[!]
\centering
\includegraphics[width=\textwidth]{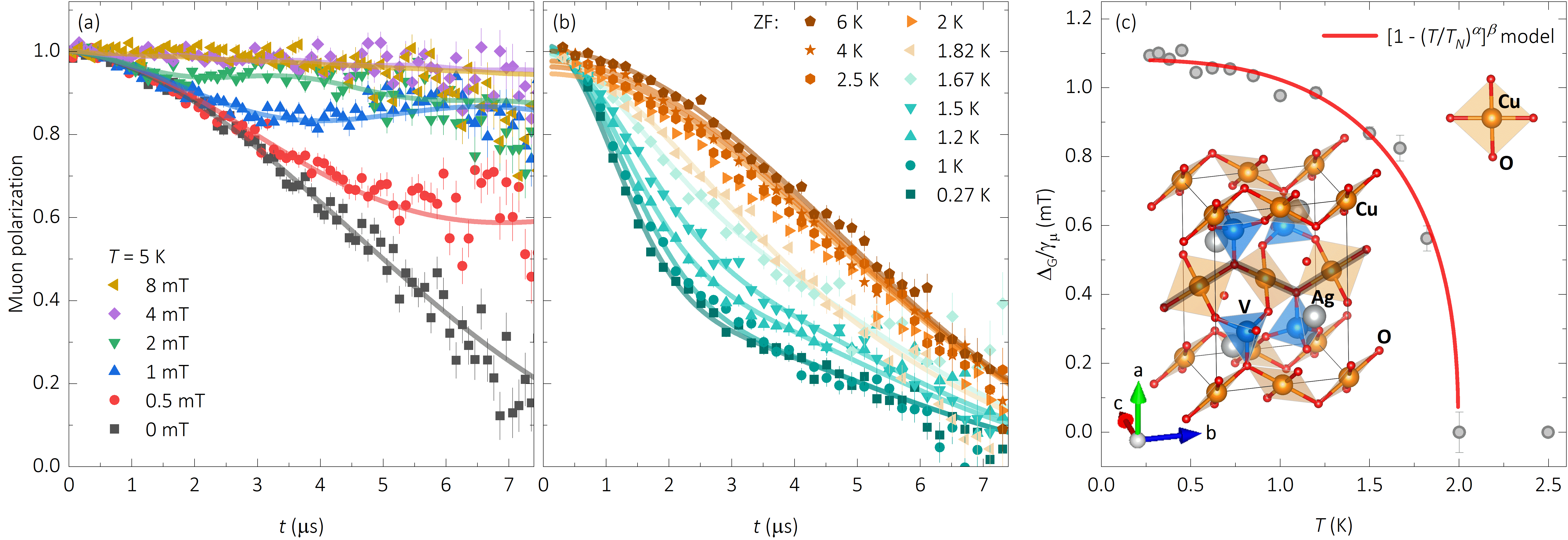}
\caption{Muon polarization decay (a) as a function of magnetic field applied along the muon polarization at 5\,K and (b) as a function of temperature in zero field (ZF). Lines in (a) correspond to the KT model for longitudinal field multiplied by an exponential relaxation, while in (b) they correspond to the sum of this KT model and a Gaussian line. (c) Extracted width of the Gaussian line as a function of temperature. 
Inset: The crystal structure of AgCuVO$_4$. The Cu$^{2+}$ spin-1/2 chains are composed of corner-sharing CuO$_4$ plaquettes (shown separately in the right corner) that run along the $b$ axis. The direction of the chains is highlighted by a thick semi-transparent black line drawn across the central chain.}
\label{fig-muSR}
\end{figure*}

AgCuVO$_4$ is an intriguing quasi-one-dimensional spin-1/2 compound, where the intrachain Cu--O--Cu bridging angle of 113$^\circ$ suggests significantly reduced antiferromagnetic intrachain interactions \cite{moller2009agcuvo}.
It crystallizes in the orthorhombic space group $Pnma$, with lattice parameters $a$\,=\,9.255(1)\,\AA, $b$\,=\,6.778(1)\,\AA, and $c$\,=\,5.401(1)\,\AA.
The Cu$^{2+}$ spin-1/2 chains, composed of corner-sharing CuO$_4$ plaquettes [see inset in Fig.\,\ref{fig-muSR}(c)], run along the $b$ axis and are separated by nonmagnetic VO$_4$ tetrahedra and Ag$^+$ ions \cite{moeller2008synthesis}.
Magnetic susceptibility shows a broad maximum near 200\,K \cite{moller2009agcuvo}, consistent with the Bonner--Fisher model \cite{bonner1964linear} for antiferromagnetic spin-1/2 chains.
The extracted exchange constant $J \approx 300$ K is supported by DFT calculations, which also reveal competing  interchain interactions of approximately $J/10$ \cite{moller2009agcuvo}.
Below 50\,K, the susceptibility increases, indicating the development of short-range correlations and/or the presence of magnetic impurities.
A small anomaly in specific heat at 2.5\,K suggests the onset of magnetic order, albeit at a temperature much lower than expected based on the size of inter- and intra-chain interactions—implying strong frustration.
However, the nature and the origin of the magnetically ordered ground state remains unclear.

In this work, we present a comprehensive investigation of the magnetic ground state of AgCuVO$_4$ using complementary local-probe muon spin spectroscopy ($\mu$SR) and neutron diffraction.
Our results reveal an incommensurate magnetic order characterized by collinear, significantly reduced magnetic moments, whose amplitude is modulated perpendicular to the spin chains.
These findings provide strong evidence for an exotic quantum state arising from magnetic frustration due to competing interchain interactions.

\section{Experimental}

The powder sample was prepared by solid state reaction of LiVO$_3$ and CuO. The powder mixture was re-ground and annealed at 395$^\circ$C several times.

Neutron powder diffraction data \cite{pregelj2021nd} were collected using two instruments: the D1B diffractometer at the Institut Laue--Langevin (France), with a neutron wavelength of $\lambda$\,=\,2.52\,\AA, and the DMC diffractometer \cite{schefer1990versatile} at the Paul Scherrer Institute (Switzerland), with $\lambda$\,=\,4.5\,\AA.
Measurements were conducted using an orange cryostat on both instruments, with additional low-temperature data on DMC obtained using a dilution refrigerator. 
Acquisition time per temperature was $\sim$1.7\,hours at D1B and $\sim$55\,hours at DMC.
Crystal and magnetic structure refinements were performed using the $FullProf$ software suite \cite{rodriguez2001fullprof}.

$\mu$SR experiments were carried out on the Dolly instrument at the Paul Scherrer Institute, equipped with a $^3$He refrigerator.
Measurements were performed in zero field (ZF) and in longitudinal magnetic fields (LF) up to 8\,mT, over a temperature range from 0.27 to 6\,K.
For ZF measurements a longitudinal muon-spin polarization, aligned along the beam direction, was chosen.

\section{Results}

\subsection{Muon spin spectroscopy}

To investigate the onset of long-range magnetic order in AgCuVO$_4$, we employ $\mu$SR, a technique highly sensitive to small internal magnetic fields and capable of distinguishing between static and dynamic magnetism \cite{yaouanc2011muon}.
At 5\,K—above the anomaly observed in the specific heat—the ZF muon polarization exhibits a monotonic, Gaussian-like decay [Fig.\,\ref{fig-muSR}(a)], indicative of a broad distribution of weak static magnetic fields, most likely of nuclear origin, as expected in the paramagnetic state \cite{yaouanc2011muon,pregelj2019elementary}.
The magnitude of the local nuclear fields depends on the muon stopping site and is therefore crucial for interpreting the $\mu$SR response in the magnetically ordered phase.
Nuclear-field-induced muon depolarization is typically described by the Kubo--Toyabe (KT) model \cite{yaouanc2011muon}, which assumes random static fields with a Gaussian distribution with a width $\Delta_\text{KT}/\gamma_\mu$, where $\gamma_\mu$\,=\,$ 2\pi\times135.5$\,MHz/T is the muon gyromagnetic ratio.
However, due to the weak nuclear magnetic fields, the characteristic minimum predicted by the KT model lies outside our experimental time window, preventing us from resolving the precise local field distribution in the ZF data.
To resolve this, we performed a LF decoupling experiment at 5\,K, applying a small external magnetic field along the initial muon spin direction.
The resulting muon polarization is well described by the KT function modified for the applied LF field \cite{yaouanc2011muon}, multiplied by a weak dynmaical exponential relaxation [Fig.\,\ref{fig-muSR}(a)].
This model yields $\Delta_\text{KT}/\gamma_\mu$\,=\,0.19(1)\,mT, consistent with typical nuclear field strengths in magnetic insulators \cite{yaouanc2011muon}.

Upon cooling, the ZF muon depolarization curve begins to change below $\sim$2.5\,K [Fig.\,\ref{fig-muSR}(b)].
Specifically, the decay becomes increasingly rapid, indicating a growth of the local static magnetic fields.
The signal retains its Gaussian character and shows no oscillations down to the lowest temperatures, suggesting that the internal magnetic fields remain broadly distributed.
The data are well described by the sum of a Gaussian function, representing the low-temperature signal, and the unmodified KT function, accounting for a residual high-temperature contribution [Fig.\,\ref{fig-muSR}(b)].

\begin{figure}[!]
\centering
\includegraphics[width=\columnwidth]{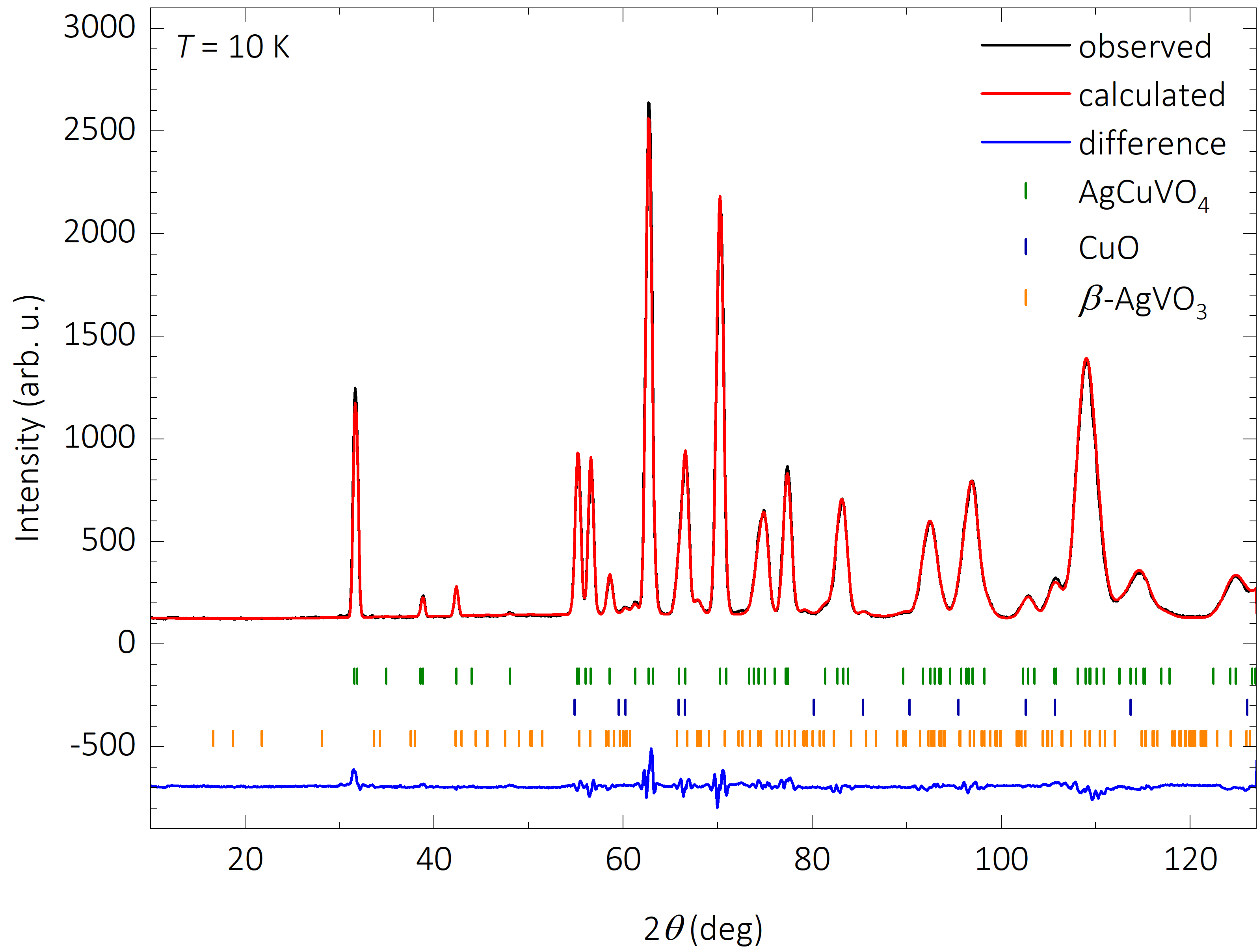}
\caption{Powder neutron diffraction pattern at 10\,K and the corresponding crystal structure refinement with impurity phases included.}
\label{fig-struc-ND}
\end{figure}

Below 2\,K, the local magnetic field—estimated from the Gaussian linewidth, $\Delta_G/\gamma_\mu$—increases rapidly and saturates below approximately  0.5\,K [Fig.\,\ref{fig-muSR}(c)].
This temperature dependence resembles that of an order parameter and is well described by the Curie--Bloch expression, $[1 - (T/T_N)^\alpha]^\beta$ \cite{marfoua2023ultra}.
Fitting the $\Delta_G/\gamma_\mu$ data between base temperature and 2\,K yields an ordering temperature $T_N$\,=\,2.0(1)\,K, a critical exponent $\beta$\,=\,0.40(8), consistent with 3D magnetic order \cite{pelissetto2002critical}.
The exponent $\alpha = 3.0(8)$ reflects the nature of the magnetic interactions and the dimensionality of the exchange lattice \cite{oguchi1960theory, liu1966nonlinear, kobler2002temperature}.
Notably, this value is close to the expected $\alpha = 5/2$ for one-dimensional half-integer spin systems \cite{kobler2002temperature}.
These findings provide clear evidence for the establishment of long-range order, characterized by a broad distribution of static internal fields at the muon site, which is more naturally expected for incommensurate rather than commensurate magnetic order.

\subsection{Neutron diffraction}

\begin{figure}[!]
\centering
\includegraphics[width=\columnwidth]{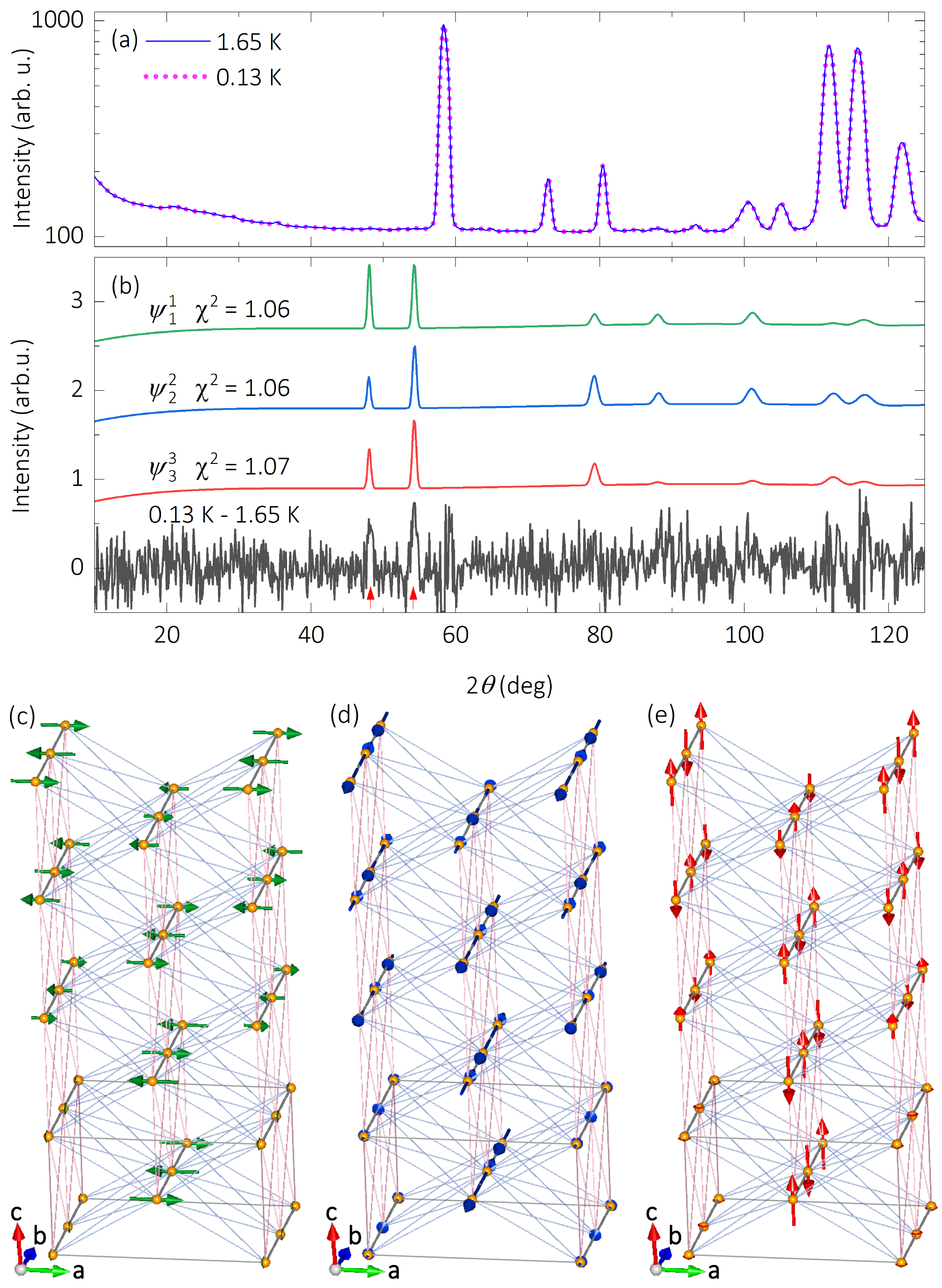}
\caption{(a) Neutron powder-diffraction pattern measured at DMC instrument at 0.13 and 1.65\,K. (b) The corresponding difference pattern (black line) and magnetic structure refinement for the three best magnetic-structure models (colored lines). The most prominent magnetic reflections are marked by arrows at the bottom. The magnetic-structure models corresponding to the (c) $\psi_1^1$, (d) $\psi_2^2$, and (e) $\psi_3^3$ basis vectors. The main intrachain interactions along the $b$ axis are marked by a dark gray line. The main interchain interactions within the $bc$ plane and those pointing out-of-plane are shown in red and blue tint, respectively. The light gray box denotes the crystal unit cell.}
\label{fig-ND-diff}
\end{figure}

\begin{table*}[!]
 \centering
 \caption{The irreducible representations (IRRs) $\Gamma_i$, $i$\,=\,1--4 and corresponding basis vectors $\psi_i^j$ for the space group $Pnma$ appearing in the magnetic representation with $\mathbf{k}_0=(0,\:0,\:0.45)$ for the magnetic site (1/2,\,0,\,1/2) of Cu atom\,1 and its three crystallographically equivalent sites: (0,\,0,\,0) of atom\,2, (1/2,\,1/2,\,1/2) of atom\,3, and (0,\,1/2,\,0) of atom\,4. The representation analysis was performed using the $BasIreps$ program incorporated in the $FullProf$ software suite \cite{rodriguez2001fullprof}. The listed basis vectors are scaled to yield normalized magnetic moments in the complex plane and $k_{0z}$ is the $z$ component of $\mathbf{k}_0$.}
  \begin{tabular*}{\linewidth}{@{\extracolsep{\fill}} c   c   c c c  c  c c c  c  c c c  c  c c c }
    \hline \hline
IRR & Basis & \multicolumn{3}{c}{Atom 1} & & \multicolumn{3}{c}{Atom 2} & & \multicolumn{3}{c}{Atom 3} & & \multicolumn{3}{c}{Atom 4} \\
      & vector & $m_x$ & $m_y$ & $m_z$ & & $m_x$ & $m_y$ & $m_z$ & & $m_x$ & $m_y$ & $m_z$ & & $m_x$ & $m_y$ & $m_z$  \\
\hline      
$\Gamma_1$ & $\psi_1^1$ & 1 & 0 & 0 & & $-e^{\pi k_{0z} i}$ & 0 & 0 & &$-1$& 0 & 0 & & $e^{\pi k_{0z} i}$ & 0 & 0  \\
           & $\psi_1^2$ & 0 & 1 & 0 & & 0 & $-e^{\pi k_{0z} i}$ & 0 & &  0 & 1 & 0 & & 0 & $-e^{\pi k_{0z} i}$ & 0 \\
           & $\psi_1^3$ & 0 & 0 & 1 & & 0 & 0 & $e^{\pi k_{0z} i}$  & & 0 & 0 &$-1$& & 0 & 0 & $-e^{\pi k_{0z} i}$ \\
$\Gamma_2$ & $\psi_2^1$ & 1 & 0 & 0 & & $-e^{\pi k_{0z} i}$ & 0 & 0 & & 1 & 0 & 0 & & $-e^{\pi k_{0z} i}$ & 0 & 0 \\
           & $\psi_2^2$ & 0 & 1 & 0 & & 0 & $-e^{\pi k_{0z} i}$ & 0 & & 0 &$-1$ & 0 & & 0 & $e^{\pi k_{0z} i}$ & 0  \\
           & $\psi_2^3$ & 0 & 0 & 1 & & 0 & 0 & $e^{\pi k_{0z} i}$  & & 0 & 0 & 1 & & 0 & 0 & $e^{\pi k_{0z} i}$  \\
$\Gamma_3$ & $\psi_3^1$ & 1 & 0 & 0 & & $e^{\pi k_{0z} i}$ & 0 & 0  & & $-1$& 0 & 0 & & $-e^{\pi k_{0z} i}$ & 0 & 0 \\
           & $\psi_3^2$ & 0 & 1 & 0 & & 0 & $e^{\pi k_{0z} i}$ & 0  & & 0 & 1 & 0 & & 0 & $e^{\pi k_{0z} i}$ & 0  \\
           & $\psi_3^3$ & 0 & 0 & 1 & & 0 & 0 & $-e^{\pi k_{0z} i}$ & & 0 & 0 &$-1$ & & 0 & 0 & $e^{\pi k_{0z} i}$  \\
$\Gamma_4$ & $\psi_4^1$ & 1 & 0 & 0 & & $e^{\pi k_{0z} i}$ & 0 & 0  & & 1 & 0 & 0 & & $e^{\pi k_{0z} i}$ & 0 & 0  \\
           & $\psi_4^2$ & 0 & 1 & 0 & & 0 & $e^{\pi k_{0z} i}$ & 0  & & 0 &$-1$ & 0 & & 0 & $-e^{\pi k_{0z} i}$ & 0 \\
           & $\psi_4^3$ & 0 & 0 & 1 & & 0 & 0 & $-e^{\pi k_{0z} i}$ & & 0 & 0 & 1 & & 0 & 0 & $-e^{\pi k_{0z} i}$ \\		   
   \hline \hline
 \end{tabular*}
 \label{tab-irreps}
\end{table*}

To determine the precise magnetic order in AgCuVO$_4$, we performed neutron powder diffraction measurements at 0.13, 1.65, and 10\,K.
The diffraction patterns exhibit sharp, resolution-limited reflections, demonstrating high crystallinity of the sample.
Rietveld refinement of the 10\,K data—taken above the magnetic ordering temperature—yields excellent agreement with the structural model [Fig.\,\ref{fig-struc-ND}].
Minor crystalline impurity phases were identified, including 0.6(1)\,wt.\% CuO and 2.3(1)\,wt.\% $\beta$-AgVO$_3$.

Cooling to 1.65\,K, just below the specific-heat anomaly, produces no discernible change in the diffraction pattern, suggesting that the ordered magnetic moments remain too small to be detected.
However, further cooling to 0.13\,K reveals two weak, but sharp, additional reflections, as seen in the difference pattern between the 0.13-K and the 1.65-K data [Fig.\,\ref{fig-ND-diff}(b)].
These magnetic reflections can be indexed with the ordering wavevector $\mathbf{k}_0=(0,\:0,\:0.45)$, indicating incommensurate magnetic modulation along the $c$ axis—perpendicular to the spin chains.
This is consistent with $\mu$SR results and with the frustrated interchain interactions [see Fig.\,\ref{fig-ND-diff}(c)-(e)] predicted by DFT calculations \cite{moller2009agcuvo}.
Notably, the magnetic reflections are also resolution-limited, confirming the long-range nature of the magnetic order.

To further characterize the magnetic order, we performed a representation analysis based on the experimentally determined magnetic propagation vector.
This analysis yields four one-dimensional irreducible representations (IRRs), each comprising three basis vectors that connect the four crystallographically equivalent Cu$^{2+}$ magnetic ions within the unit cell (Table\,\ref{tab-irreps}).
These basis vectors were used as input for magnetic structure refinement against the difference pattern between the 0.13-K and the 1.65-K neutron diffraction data.

Out of the twelve possible basis vectors, only three—$\psi_1^1$, $\psi_2^2$, and $\psi_3^3$—are consistent with the experimental data, yielding comparable refinement qualities [Fig.\,\ref{fig-ND-diff}(b)].
This is to be expected, as all three magnetic structure models [Fig.\,\ref{fig-ND-diff}(c)--(d)] share some key features:
(i) antiferromagnetic alignment of neighboring spins along the $b$ axis, consistent with dominant antiferromagnetic intrachain interaction;
(ii) a phase shift of $-e^{\pi k_{0z} i}$ between atoms 1 and 2, as well as 3 and 4, suggesting competing magnetic interactions along the $c$ axis (Table\,\ref{tab-irreps}); and
(iii) a collinear, amplitude-modulated magnetic structure with a magnetic moment amplitude of 0.13(2)\,$\mu_\mathrm{B}$, where $\mu_\mathrm{B}$ is the Bohr magneton.

This significantly reduced ordered moment, compared to the expected full Cu$^{2+}$ ($S$\,=\,1/2) moment of $\sim$1\,$\mu_\mathrm{B}$, is consistent with the quasi-one-dimensional and frustrated nature of the system, which enhances quantum fluctuations and entanglement \cite{schulz1996dynamics, moller2009agcuvo}.
The three models differ only in the orientation of the magnetic moments—$\psi_1^1$ along $a$, $\psi_2^2$ along $b$, and $\psi_3^3$ along $c$—which cannot be distinguished based on the current powder neutron diffraction data.

\section{Discussion}

\begin{figure}[!]
\centering
\includegraphics[width=\columnwidth]{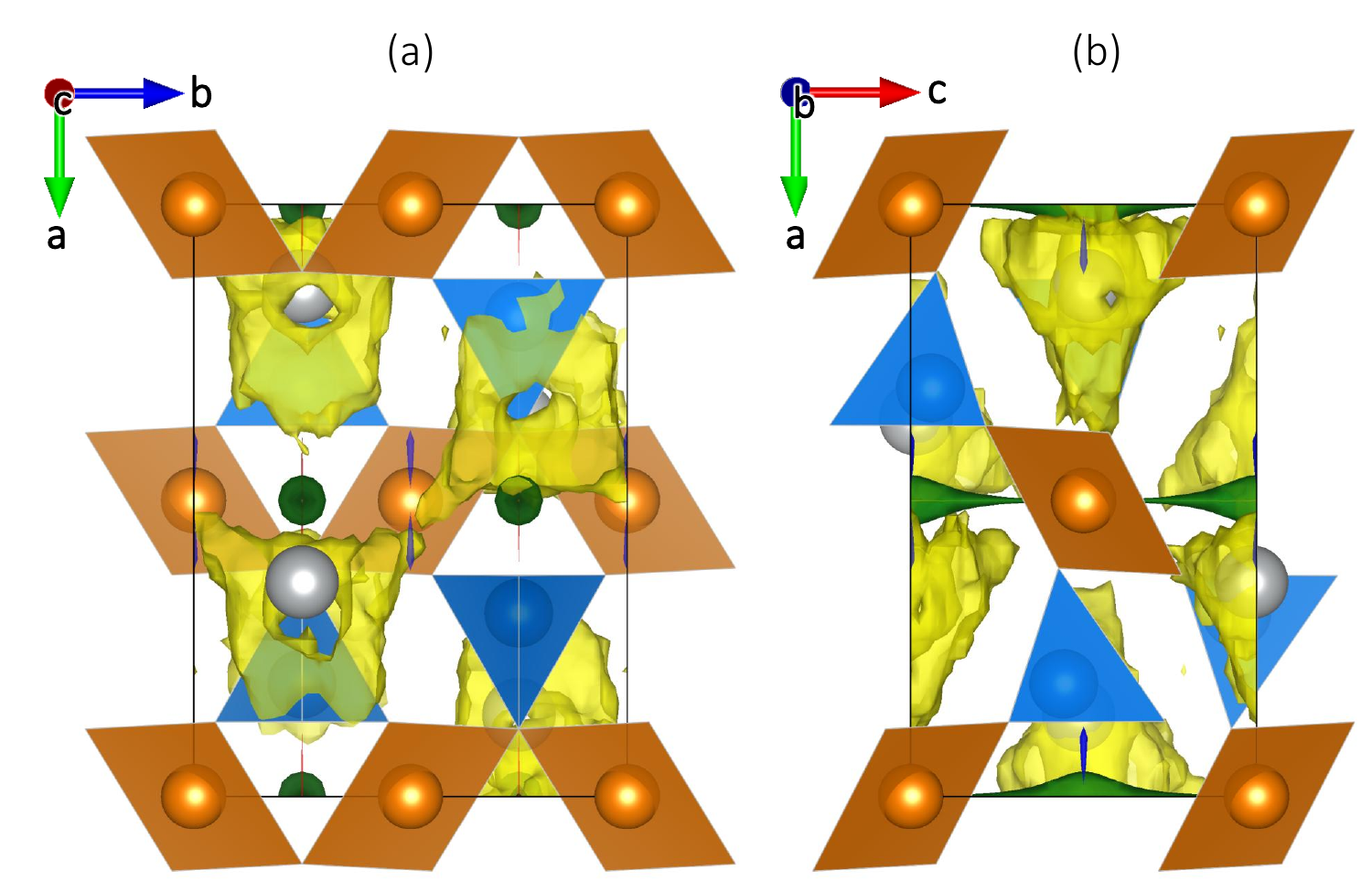}
\caption{The crystal structure of AgCuVO$_4$ for two different orientations (a) and (b), respectively. The atoms are represented by colored spheres, i.e., Cu (brown), V (blue), and Ag (light gray), while O atoms positioned at the corners of CuO$_4$ (brown) and VO$_3$ (green) units are skipped for clarity. The light yellow color represent the area, where the mean nuclear magnetic field equals 0.35\,mT, while other colors represent areas, where magnetic field equals 1.1\,mT for $\psi_1^1$ (green), for $\psi_2^2$ (blue), and for $\psi_3^3$ (red). The latter two are very narrow and almost too small to be seen.}
\label{fig-struc-stop}
\end{figure}

To further refine the magnetic structure model, we combined the $\mu$SR results with the neutron diffraction data.
We first calculated the mean magnetic field arising from the randomly oriented nuclear magnetic moments throughout the unit cell, which corresponds to $\sqrt{2}\Delta_\text{KT}/\gamma_\mu$, as defined by the KT model \cite{yaouanc2011muon}.
The minimum calculated mean-field value is 0.24\,mT, closely matching the experimentally determined value of 0.27(3)\,mT at 5\,K.
Regions where the nuclear field is close to the experimental value (for instance, below 0.35\,mT) are located primarily around the Ag$^+$ ions [Fig.\,\ref{fig-struc-stop}] and represent the most probable muon stopping sites.

Given the spatial localization of these regions, different magnetic structures may produce significantly different local magnetic fields within them.
We therefore calculated the maximum electronic magnetic fields generated by the three candidate magnetic structures identified via neutron diffraction.
In Fig.\,\ref{fig-struc-stop}, we overlay the regions where the electronic magnetic field equals $\Delta_G/\gamma_\mu$\,=\,1.1\,mT—corresponding to the width of the local field distiribution at the muon site at base temperature [Fig.\,\ref{fig-muSR}]—onto the potential muon stopping region.
The overlap between these two regions indicates the most likely muon stopping position, consistent with the $\mu$SR response both above and below the magnetic transition.

This analysis reveals that such overlap occurs only for the $\psi_1^1$ and $\psi_2^2$ models, while the fields generated by the $\psi_3^3$ model are too large.
These findings strongly support that the magnetic order in AgCuVO$_4$ is described by either $\psi_1^1$ or $\psi_2^2$, corresponding to antiferromagnetic spin chains with collinear magnetic moments oriented along the $a$ or $b$ axis, respectively, and amplitude modulated along the $c$ axis [Fig.\,\ref{fig-ND-diff}(c),(d)].
This configuration is consistent with the presence of frustrated interchain interactions that compete among their selves and with the main intrachain interaction, as they form triangular motifes within the spin lattice [Fig.\,\ref{fig-ND-diff}(c)-(d)].

\section{Conclusions}

Using muon spin spectroscopy and neutron diffraction, we identified two closely related candidate incommensurate magnetic structures for the ground state of AgCuVO$_4$.
Both structures comprise antiferromagnetic spin-1/2 chains with collinear magnetic moments oriented along either the $a$ or $b$ axis.
These moments exhibit a a significantly reduced maximum amplitude of 0.13(3)\,$\mu_\mathrm{B}$ due to quantum entanglement.
The amplitude is modulated along the $c$ axis—perpendicular to the chains, which run along the $b$ axis—indicating the presence of frustrated interchain interactions.
Our findings establish AgCuVO$_4$ as an excellent model system for exploring the effects of geometrical frustration arising from exchange couplings between antiferromagnetic quantum spin chains.
The dynamical response of this system presents a promising avenue for future exploration, with the potential to provide deeper insight into the interplay between quantum entanglement and geometrical frustration.

\begin{acknowledgments}

This work has been funded by the Slovenian Research Agency (projects No. J2-2513, J1-50012, and N1-0356, and program No. P1-0125).
This work is partly based on experiments performed at the Swiss spallation neutron source SINQ, Paul Scherrer Institute, Villigen, Switzerland.

\end{acknowledgments}

\end{document}